\documentclass[12pt]{iopart}

\usepackage{graphicx}
\usepackage{color}
\usepackage{iopams}

\begin{document}

\title[Benchmark Quantum Monte Carlo calculations]{Benchmark Quantum Monte Carlo calculations of the ground-state kinetic, interaction, and 
total energy of the three-dimensional electron gas}
\author{I~G~Gurtubay$^{1,2}$, R Gaudoin$^{2}$ and
J~M~Pitarke$^{1,3}$}
\address{
$^{1}$ Materia Kondentsatuaren Fisika Saila, Zientzia eta Teknologia Fakultatea,
Euskal Herriko Unibertsitatea, 644 Posta kutxatila, E-48080 Bilbo,
Basque Country, Spain}
\address{$^{2} $Donostia International Physics Center (DIPC),
Manuel de Lardizabal Pasealekua, E-20018 Donostia, Basque Country, Spain}
\address{$^{3}$ CIC nanoGUNE Consolider and Centro de  F\'{\i}sica Materiales
CSIC-UPV/EHU, Tolosa Hiribidea 76, E-20018 Donostia, Basque Country, Spain}
\ead{idoia.gurtubay@ehu.es}

\begin{abstract}
We report variational and diffusion Quantum Monte Carlo
ground-state energies of the three-dimensional electron gas
using a model periodic Coulomb interaction and backflow corrections
for N=54, 102, 178, and 226 electrons. We remove finite-size effects by 
extrapolation and we find lower energies than previously reported.
Using the Hellman-Feynman operator sampling method introduced
in Phys. Rev. Lett. {\bf 99}, 126406 (2007), we compute accurately,
 within the fixed-node approximation, the separate kinetic and interaction
contributions to the total ground-state energy. 
The difference between the interaction energies obtained
from the original Slater-determinant nodes and
 the backflow-displaced nodes is found to be
considerably larger than the difference between
 the corresponding kinetic energies.
\end{abstract}

\pacs{71.10.Ca, 71.15.-m}
\submitto{JPCM}

\maketitle

\section{\label{sarrera}INTRODUCTION}
The homogeneous electron gas (HEG), which represents the simplest possible prototype of a many-fermion system, has been over the years a topic of intense research, as
it often provides a good approximation for the description of valence electrons in simple metals and represents the basic ingredient for local and semilocal density-functional approximations.

One of the first exhaustive calculations of the ground-state energy of an interacting three-dimensional (3D) HEG was performed by Ceperley \cite{Ceperley-1978},
 using stochastic numerical methods. In this work, Ceperley used variational Monte Carlo (VMC) to obtain an upper bound to the ground-state energy.
 More accurate ground-state energies can be computed by using the more
 sophisticated diffusion Monte Carlo (DMC) approach, which projects out the
 true ground state of the system from a trial wave function 
\cite{Foulkes_rmp-2001}. However, this method yields a bosonic ground state
 even for a fermionic system which needs to have an antisymmetric ground state.
 In order to overcome this problem, 
the fixed-node (FN) approximation \cite{Anderson-1976} 
 has been applied; this approximation  
constrains the nodes of the
ground-state wave function to those of a trial wave function.
In spite of this constraint, the FN method has proven to be useful for
 the calculation of the ground-state energy and other electronic properties
for atoms \cite{Ma_noble-2005,Drummond_etal-Ne-2006},
molecules \cite{Gurtubay_etal-2006,Gurtubay-Needs-2007},
solids \cite{Maezono_etal-2007,Sola-et-al-2009} and the two-dimensional (2D)
electron gas \cite{Tanatar-Ceperley-1989, Attaccalite-2D-2002, Drummond-Needs-2dHEG-2009}.
 As an alternative to overcome the sign-problem for fermions, Ceperley and
Alder \cite{Ceperley-Alder-PRL-1980} developed the so-called
released-node DMC, which has the limitation that statistical fluctuations
 grow very rapidly and statistical noise can dominate the signal
 even before converging to the ground state.
 The size of the system that can be simulated is also limited in this method.
 Nonetheless, these released-node data have been widely used in the framework of
 density-functional calculations.

With the improvement of computing capabilities and algorithms, 
new calculations have been attempted in order to improve the Ceperley-Alder DMC data. 
Ortiz and Ballone \cite{Ortiz-Ballone-1994}
extended these calculations to larger system sizes and polarized systems. 
Kwon {\it et al.} \cite{kwon98:_effec} introduced backflow and three-body 
correlations in the wave function, showing an improvement in the
FN result beyond that given by Slater-Jastrow wave functions for a given
 finite system
({see also \cite{lopez-rios_etal-2006}, \cite{Zong_etal-2002}
and \cite{Holzmann-93}}).
Both Ortiz and Ballone \cite{Ortiz-Ballone-1994} and
 Kwon {\it et al.} \cite{kwon98:_effec} 
  studied the size dependence and extrapolated their results to the 
thermodynamic limit; however, they assumed that the size dependence
for VMC and DMC is the same, and they used VMC data to extrapolate the corresponding DMC calculations.

Fixed-node DMC is known to yield the exact ground-state energy of a 
many-electron system for a given nodal structure.
 Nevertheless, the DMC expectation value of any operator 
that does not commute with 
the Hamiltonian differs from the exact value, the error being linear in the difference between the trial and the projected wave function. Recently, an effective method based on the Hellman-Feynman (HF) theorem
was devised to calculate the exact expectation value
of such an operator \cite{Gaudoin-Pitarke-HFS} as, for example, the interaction energy.

In this paper, we use the HF sampling introduced in \cite{Gaudoin-Pitarke-HFS} to
 report benchmark DMC calculations  of the two 
(kinetic and interaction) separate contributions to the ground-state energy of 
a paramagnetic 3D electron gas with 
$r_s=2.$\footnote{$r_s$ is the dimensionless parameter $r_s= a/a_0$ where $a_0$
is the Bohr radius and $a$ is the radius of a sphere that encloses one
electron on average.} 
We include backflow correlation effects
and we demonstrate that previously extrapolated results are artificially lowered
by assuming that the VMC and DMC size dependences coincide. 
However, we find that  DMC size dependences are similar with or without backflow
 correlation effects. 
The use of the HF sampling allows to compute the exact 
(within the fixed node approximation)
 interaction energy  in the framework of the modified periodic Coulomb (MPC) 
scheme, and we show that this interaction energy can also be obtained 
from the integration of the spherically averaged wave-vector dependent 
diagonal structure factor  \cite{Gaudoin_etal-SFU-dmc-2009}. 
  Furthermore, we find that the HF-sampling scheme works efficiently even for very large systems.

Hartree atomic units (a.u.) are used throughout, i.e.,
 $\hbar=|e|=m_e=4\pi\epsilon_0$=1. 
The atomic unit of energy is $e^2/a_0=27.2~{\rm eV}$, $a_0$ being the Bohr
 radius.
 All the calculations presented in this work have been performed by using the
 {\textsc {casino}} code \cite{casinomanual-2.1}.

\section{\label{qmc}QUANTUM MONTE CARLO}

In VMC the ground-state energy, $E_{V\!M\!C}$,
is estimated as the expectation value
of the Hamiltonian with an approximate trial wave function, $\Psi_T$: 
$E_{V\!M\!C}~=~\langle\hat H\rangle_{V\!M\!C}~=~{\langle\Psi_T|\hat H|\Psi_T\rangle}/
{\langle\Psi_T|\Psi_T\rangle}$. 
The integrals are evaluated by importance-sampled Monte Carlo 
integration. The trial wave function contains parameters,
whose values are obtained from an optimization procedure formulated
within VMC.  There are no restrictions on the form of the trial wave
function, and VMC does not suffer from a fermion
sign problem.  However, the choice of the approximate trial wave
function is very important, as it directly determines the accuracy
of the calculation. We have used VMC methods mainly to optimize the parameters involved in the trial wave functions by variance and energy minimization; our most accurate calculations, however, have been performed within DMC.

In DMC the ground-state component of a trial wave function is
projected out by evolving an ensemble of electronic configurations
using the imaginary-time Schr\"odinger equation.  The fermionic
symmetry is maintained by the fixed-node
approximation \cite{Anderson-1976},  in which the nodal surface of the DMC wave
function is constrained to equal that of the trial wave function. 
The fixed-node DMC energy,
$E_{F\!N}~=~\langle\hat H\rangle_{F\!N}~=~{\langle\Psi_0^{F\!N}|\hat H|\Psi_0^{F\!N}\rangle}
/{\langle\Psi_0^{F\!N}|\Psi_0^{F\!N}\rangle}$,
is higher than the exact ground-state energy
$\langle\hat H\rangle~=~{\langle\Psi_0|\hat H|\Psi_0\rangle}
/{\langle\Psi_0|\Psi_0\rangle}$,
and they become equal only when the fixed nodal surface is that of the exact ground
state, $\Psi_0$. Apart from the fixed-node error, DMC yields the true ground-state energy independently of the form chosen for the trial wave function. The fixed-node error can be reduced by optimizing the nodal surfaces of the trial wave function.

\subsection{The Slater-Jastrow backflow trial wave function}

The standard Slater-Jastrow (SJ) wave function can be written as
\begin{equation}\label{eq:psiSJ}
\Psi_{\rm SJ}({\bf R}) = e^{J({\bf R})} \Psi_{\rm S}({\bf R}),
\end{equation}
where {\bf R} is a 3$N$-dimensional vector denoting the position ${\bf
r}_i$ of each electron. The nodes of $\Psi_{\rm SJ}({\bf R})$ are
defined by the Slater part of the wave function, $\Psi_{\rm S}({\bf
R})$, which takes the form
$\Psi_{\rm S}\,=\,D_{\uparrow}\,D_{\downarrow}$, $D_{\sigma}$ being a Slater
determinant of single-particle orbitals of spin $\sigma$.
In the case of a HEG, these orbitals are Hartree-Fock solutions
for the finite periodically repeated electron gas, which are simply plane waves.
The number of electrons $N$ has been chosen such that the ground state
 is a closed shell configuration,
so the wave function can be chosen to be real and there is no degeneracy.
The Jastrow correlation factor, $e^{J({\bf R})}$, contains an
electron-electron and a plane-wave term,
 as described in \cite{Drummond_Towler_Needs-2004}.
We did not include a symmetric three-electron Jastrow term.\footnote{We
found that for the density considered in this work, the effect
of the three-body term was not statistically significant in conjunction
with backflow, which gives the best energy. This is in agreement with 
\cite{lopez-rios_etal-2006}.}
The Jastrow factor, being a positive definite function,
 keeps electrons away from each other and greatly
improves wave functions in general, but it does not modify the nodal
surface of the wave function.

One way of reducing the FN error is to alter the nodes of the wave
function by introducing backflow correlations \cite{lopez-rios_etal-2006}, 
thus replacing the coordinates {\bf R} in the
Slater part of the wave function by the collective coordinates {\bf
X}. The Slater-Jastrow backflow (SJB) trial wave function reads
\begin{equation}\label{eq:psiiBF}
\Psi_{\rm SJB}({\bf R}) = e^{J({\bf R})} \Psi_{\rm S}({\bf X}).
\end{equation}
The new coordinates for each electron are given by
\begin{equation}\label{eq:psi}
{\bf x}_i = {\bf r}_i+\xi_i({\bf R}),
\end{equation}
$\xi_i$ being the backflow displacement of particle $i$, which depends
on the position of every electron in the system.  
Details of the specific form of the backflow function  used for the HEG
can be found in \cite{lopez-rios_etal-2006}.

\subsection{Hellman-Feynman sampling}
\label{HFSmethod}

Given an arbitrary operator $\hat O$, 
the fixed-node DMC method yields by construction the normalized expectation value
$\langle\hat O\rangle_{F\!N\!-\!D\!M\!C}~=~{\langle\Psi_T|\hat O|\Psi_0^{F\!N}\rangle}
/{\langle\Psi_T|\Psi_0^{F\!N}\rangle}$,
which is not the true FN ground-state expectation value 
$\langle\hat O\rangle_{F\!N}~=~{\langle\Psi_0^{F\!N}|\hat O|\Psi_0^{F\!N}\rangle}
/{\langle\Psi_0^{F\!N}|\Psi_0^{F\!N}\rangle}$, 
unless the operator $\hat O$ commutes with the Hamiltonian,
the leading term of this error being linear in the difference between $\Psi_T$ and $\Psi_0^{F\!N}$.
In conjuntion with VMC, this error can be reduced by one order by using
the so-called extrapolated estimator,  
 $2\langle\hat O\rangle_{F\!N\!-\!D\!M\!C}-\langle\hat O\rangle_{V\!M\!C}$.
In practice, extrapolated estimators work well when the trial wave function
is very close to the ground state, but they can be untrustworthy
when the trial wave function is poor. A correct sampling for these operators 
can be achieved by using future walking \cite{Kalos-66,barnett_etal-1991} or
reptation Monte-Carlo \cite{Baroni-Moroni-1999};
however, these methods aim at sampling $\Psi_0^{F\!N} \Psi_0^{F\!N}$ instead of the 
regular DMC distribution \cite{Foulkes_rmp-2001}, $\Psi_T\Psi_0^{F\!N}$,
so they are not straightforward additions to the DMC algorithm.

An alternative to achieve a correct sampling of operators
 (diagonal in real space) that do not commute with the Hamiltonian 
has been reported recently \cite{Gaudoin-Pitarke-HFS}. The advantage of this
so-called HF sampling
is that it samples the usual DMC distribution,
$\Psi_T \Psi_0^{F\!N}$, and it is, therefore, straightforward to implement on a DMC algorithm.
We give here a sketch of the method; a detailed derivation can
be found in \cite{Gaudoin-Pitarke-HFS}.
Given a Hamiltonian of the form $\hat H(\alpha) = \hat H + \alpha \hat O$ and
the associated fixed-node ground-state energy
$E_{F\!N}(\alpha)=\langle\hat H(\alpha)\rangle_{F\!N}$, 
first-order perturbation theory for $\Psi_0^{F\!N}$ yields a fixed-node
equivalent of the HF theorem\footnote{The Hellman-Feynman theorem has also been applied previously in QMC for deriving accurate expectation values of observables
\cite{Assaraf-Caffarel-2003}.}
\begin{equation}
\langle\hat O\rangle_{F\!N}=\frac{\partial E_{F\!N}( \alpha )}{\partial \alpha} \Big|_{\alpha = 0}.
\end{equation}
Direct application of the HF derivative to the regular
DMC algorithm at timestep $i$ gives:
\begin{equation}
O^E_i=\frac{\partial E_i( \alpha )}{\partial \alpha} \Big|_{\alpha = 0}=
{\overline{O^L_i}}-t \left( {\overline{E^L_i X_i}} - 
{\overline{E^L_i}} \cdot {\overline{X_i}} \right),
\end{equation}
$t$ is an auxiliary parameter and 
${\overline{O^L_i}}$ is the standard DMC estimator at time step $i$: 
\begin{equation}\label{eqOLi}
{\overline{O^L_i}}= \sum_{j}^{N_w} \omega_{i,j} \,\,O^L_{i,j},
\end{equation}
where $\omega_{i,j}$ is the total weight
of walker $j$,  
and  $O^L_{i,j}=\hat O \Psi_T/\Psi_T$ with the trial wave
function, $\Psi_T$, evaluated for walker $j$  at time step $i$.
$\overline {X_i}$ is the DMC estimator at time step $i$ 
of a new variable per operator
$X_{i,j}~=~\frac{1} {i} \displaystyle\sum_{k=1}^{i} O_{k,j}^L $.
The fixed-node DMC estimate, which we call
 $\langle\hat O\rangle_{F\!N\!-\!D\!M\!C}$, is obtained
averaging {\Eref{eqOLi}} over all $i$.

The correction term, $\Delta O^E_i = -t \left( {\overline{E^L_i X_i}} - 
{\overline{E^L_i}} \cdot {\overline{X_i}} \right)$, involves
information which can be directly obtained from the DMC algorithm, such
as the local energy, 
$E^L_{i,j}=\hat H \Psi_T/\Psi_T$,  and the new variable,
$\overline {X_i}$, which 
involves no more than an extra summation step during the sampling.
The true FN estimate, which we call $\langle\hat O\rangle_{H\!F}$, 
is obtained averaging $O^E_i$ over all $i$.

Exponentially limiting the depth of the history in $X_{i,j}$ allows  one
to considerably improve sampling and reduce statistical noise
without reintroducing a significant bias \cite{Gaudoin-errors-2009}.

\section{Details of the calculations}

We have studied an unpolarized 3D HEG consisting of
54, 102, 178, and 226 electrons in a face-centered-cubic simulation
cell subject to periodic boundary conditions.

The parameters in the SJB trial wave functions were 
obtained by first minimizing the variance of the local 
energy \cite{Kent_Needs_Rajagopal-1999,Drummond_Needs-2005} and then 
minimizing the energy \cite{Brown-thesis-2007}.
The specific forms of the Jastrow and backflow functions are described in 
\cite{Drummond_Towler_Needs-2004} and \cite{lopez-rios_etal-2006}.
The 2-body Jastrow term ($U$) and  backflow ($\eta$) terms  consist on power expansions in the electron-electron distance with expansion orders $N_U$=$N_{\eta}$=8 and the 
parameters were allowed to depend on the spin parameters of the electron pairs.
The cutoff lengths at which both $U$ and $\eta$ go smoothly to zero, $L_U$ and $L_{\eta}$,  were optimized, but they adjusted themselves to the maximum allowed values, i.e., the Wigner-Seitz radius.
The plane-wave term in the Jastrow factor included 128
reciprocal-lattice vectors of the simulation cell.
We used a sufficiently small time step (0.003~a.u.), to avoid finite-time-step
errors, and a target population of 800   configurations in all our DMC calculations,
making population-control bias negligible.

Because our QMC calculations are performed using a finite simulation cell subject
to periodic boundary conditions,\footnote{We used the Gamma point only. 
The recent analysis of
Drummond {\it et al.} \cite{Drummond_etal-FSE-2008} showed that
 the $\Gamma$-point extrapolated SJ-DMC energies of a HEG agree
 within error bars with the results obtained by using the more
sophisticated twist-averaged boundary conditions \cite{Lin_et_al-TABC-2001}.
}
 the energy per particle is calculated at several
system sizes and then the results are extrapolated to infinite system size. 
Finite-size effects in the kinetic energy are typically taken into account 
by noting that they are roughly proportional to the corresponding finite-size errors
in the Hartree-Fock kinetic energy.
Coulomb finite-size effects in the interaction energy of a HEG,
 which arise from the spurious interaction
between an electron and the  periodically repeated  copies of its 
exchange-correlation (xc) hole, 
 can be reduced either by  adding the correction  proposed
by Chiesa {\it et al.} \cite{Chiesa-FSE} to the usual Ewald energy 
or by using the MPC \cite{ Drummond_etal-FSE-2008, Fraser_etal-96, Williamson_etal-97, Kent_etal-1999}
 interaction.
In this work, Coulomb finite-size effects were reduced by using the MPC scheme,
where  the Hartree energy is calculated with the Ewald
interaction while the xc energy is calculated using $1/r$
within the minimum image convention, that is, reducing the
interelectron distance into the Wigner-Seitz cell of the simulation cell.
The MPC interaction was used for the branching factors and for computing
energies in DMC. According to Drummond et al.  \cite{ Drummond_etal-FSE-2008}
the Ewald interaction distorts less the xc hole when calculating branching
factors in DMC. However MPC branching is
 faster and so potentially allows bigger systems.
The trade-off is therefore between better convergence due to the xc hole being
more accurate, or due to being able to go to bigger systems quicker.
 We have chosen the latter.

\section{Results and discussion}
\subsection{Ground-state total energy} \label{totalE}

\begin{figure}[htb]
\includegraphics*[width=0.95\linewidth]{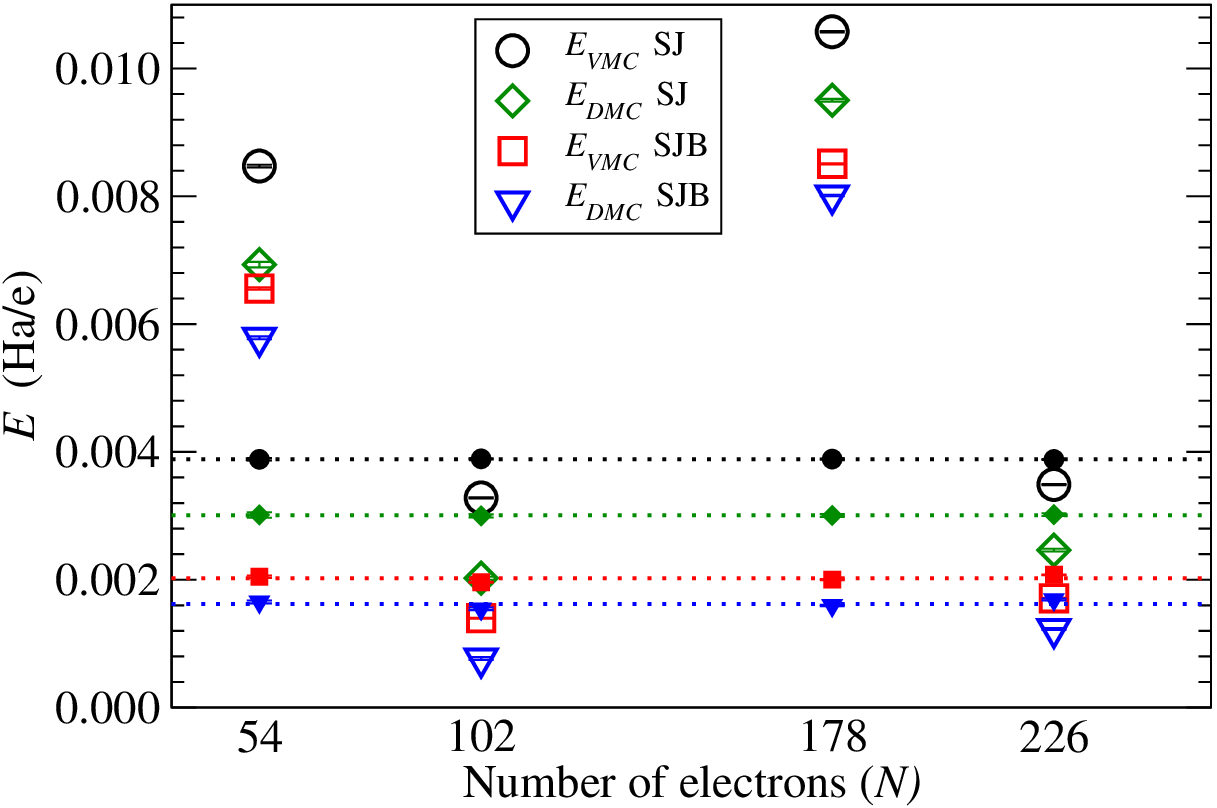} 
\caption{\label{figE-N}
(Color online) Big open symbols: VMC and DMC ground-state total energies,
$E_{V\!M\!C}$ and $E_{D\!M\!C}$, from table~\ref{tableE-N}. Horizontal dotted
 lines: The extrapolated values $E_\infty$
 (also quoted in table~\ref{tableE-N}), as obtained from \Eref{eqfit2D}.
 Small filled symbols: The extrapolated values,
$E_N-b_1\, \Delta T_{\rm HF}(N) -\frac{b_2}{N}$,
that we have obtained from \Eref{eqfit2D} for each $N$ and fixed values of $b_1$ and $b_2$.}
\end{figure}

VMC and DMC ground-state total energies of a HEG of $r_s=2$ for
$N$=54, 102, 178, and 226 electrons, as obtained by using either SJ or SJB trial wave functions within the MPC scheme, are given in table~\ref{tableE-N}. 
These energies are also displayed by open symbols in \fref{figE-N} with their
 corresponding error bars, which are less than 20\% of the symbol size.
 Both VMC and DMC ground-state energies show the usual finite-size effects. 
For comparison, table~\ref{tableE-N} also displays the DMC-SJ energies obtained by
using the Ewald interaction.

\begin{table}[h!]
\caption{\label{tableE-N}
VMC and DMC ground-state total energies, $E_{V\!M\!C}$ and $E_{D\!M\!C}$, as obtained with the use of SJ and SJB trial wave functions for a HEG of $r_s$=2 for $N$= 54, 102, 178, and 226 electrons and the MPC interaction. The column denoted with Ewald
shows the DMC energies for the SJ wave function and the Ewald interaction.
 The entry $N=\infty$
corresponds to the extrapolated values obtained from \Eref{eqfit2D}.}
\begin{indented}
\lineup
\item[]
\begin{tabular}{@{}l l l l  l l}
\br
 & \multicolumn{3}{c}{SJ wave function} & \multicolumn{2}{c}{SJB wave function} \\
& \multicolumn{2}{c}{MPC} & \multicolumn{1}{c}{Ewald} & \multicolumn{2}{c}{MPC}\\
$N$ & \multicolumn{1}{c}{$E_{V\!M\!C}$ (Ha/e)} &
\multicolumn{1}{c}{$E_{D\!M\!C}$  (Ha/e)} & 
\multicolumn{1}{c}{$E_{D\!M\!C}$  (Ha/e)} &
\multicolumn{1}{c}{$E_{V\!M\!C}$ (Ha/e)}  &
 \multicolumn{1}{c}{$E_{D\!M\!C}$ (Ha/e)}\\
\mr
54 & 0.00847(2)& 0.00693(4) & 0.00426(2) & 0.00656(2) & 0.00579(2) \\
102& 0.00328(1)& 0.00202(2) & 0.00081(2) & 0.001398(8)& 0.00077(2)\\
178& 0.01057(1)& 0.00950(2) & 0.00892(2) & 0.008510(8)& 0.00801(1)\\
226& 0.003490(8)&0.00246(2) & 0.00202(2) & 0.001710(5)& 0.00123(1)\\
\hline
$\infty$  & 0.003886(5)  &  0.00301(1)& 0.00331(3) & 0.002021(4) & 0.001621(7)\\
\br
\end{tabular}
\end{indented}
\end{table}

The issue of finite-size corrections to the kinetic and interaction contributions to the ground-state energy has been adressed by Ceperley and 
co-workers \cite{Ceperley-1978,Ceperley-Alder-1987,Tanatar-Ceperley-1989,
kwon98:_effec}. They proposed separate extrapolation terms for
the kinetic and interaction contributions using the following
extrapolation formula:
\begin{equation}\label{eqfit2D}
E_N=E_\infty+b_1\, \Delta T_{\rm HF}(N) +\frac{b_2}{N}, 
\end{equation}
where $E_\infty$, $b_1$, and $b_2$ are parameters to be fitted.
$\Delta T_{\rm HF}(N)$ is the difference between the Hartree-Fock kinetic energies of the finite and infinite systems, 
$\Delta T_{\rm HF}(N)~=~T_{\rm HF}(N)~-~T_{\rm HF}({\infty})$,
and the term $b_2/{N}$ accounts for the finite-size effects arising in the
interaction energy. Recently it has been shown that the  $b_2/{N}$ term
also corrects for the neglect of  long-range correlation effects 
in the kinetic energy \cite{Chiesa-FSE}.

Ortiz and Ballone \cite{Ortiz-Ballone-1994} considered an extrapolation
of the form
\begin{equation}\label{eqfitOB}
 E_N=E_\infty+ \Delta T_{\rm HF}(N) -\left(\frac{N}{b_0}-
 \frac{1}{\Delta v_{\rm HF}(N) }
 \right)^{-1},
 \end{equation}
with only 2 parameters to be fitted,  $E_\infty$ and $b_0$.
$\Delta v_{\rm HF}(N) $ in \Eref{eqfitOB} is the difference between the Hartree-Fock exchange energies of the finite and infinite system,
$\Delta v_{\rm HF}(N) =v_{\rm HF}(N) -v_{\rm HF}( {\infty})$.
\footnote{Since the exchange hole entering the Hartree-Fock exchange energy
 is very long ranged compared to the exchange-correlation hole, the
finite-size correction $\Delta v_{\rm HF}(N)$ entering \Eref{eqfitOB}
 might not be appropriate in an extrapolation scheme for QMC energies.}
 \Eref{eqfit2D} and \eref{eqfitOB} were found to yield similar 
results. We have found, however, that \Eref{eqfit2D} yields in all cases
better fits of the QMC data, so that all our extrapolations have been
 carried out by
using \Eref{eqfit2D}.\footnote{The adjusted R-squared values for the fits are: 
VMC-SJ: 0.999995, VMC-SJB: 0.999395, DMC-SJ: 0.999004, and DMC-SJB: 0.999983.
} 
The horizonal dotted lines of \fref{figE-N} display the extrapolated
 values $E_\infty$ (also quoted in table~\ref{tableE-N}) that we have found
 from \Eref{eqfit2D}. These extrapolated values indicate that the optimization of the nodes of the trial wave function (which is achieved by replacing the SJ
 trial wave function by the SJ-backflow trial wave function) lowers
 the ground-state energy
 considerably:\footnote{This corroborates the result reported in 
\cite{lopez-rios_etal-2006}  to a small system of 54 electrons.}
 2~mHa/e at the VMC level and 1.4~mHa/e at the DMC level; moreover, the optimized VMC-SJB ground-state energy happens to be 1~mHa/e lower than its fixed-node counterpart DMC-SJ.

\begin{table}[ht!]
\caption{\label{tableEinf-comp}
Top: The extrapolated VMC and DMC ground-state total energies
$E_\infty$ of table~\ref{tableE-N}, as obtained with the use of SJ and SJB trial wave functions for a HEG of $r_s=2$.
Middle: The VMC calculation of Cerperley (C-VMC) \cite{Ceperley-1978}, 
the released-node DMC calculation of Ceperley and Alder (CA-DMC-RN)
\cite{Ceperley-Alder-PRL-1980}, the fixed-node VMC and DMC calculations of
Ortiz and Ballone (OB-VMC-SJ and OB-DM-SJ) \cite{Ortiz-Ballone-1994}
 and Kwon {\it et al.} (KCM-VMC-SJ and KCM-DMC-SJ) \cite{kwon98:_effec},
and the backflow DMC calculation of Kwon {\it et al.} (KCM-DMC-SJB)
\cite{kwon98:_effec}.
Bottom: DMC ground-state total energies, but now obtained by using 
(as in \cite{Ortiz-Ballone-1994} and \cite{kwon98:_effec})
the VMC parameters $b_1$ and $b_2$
(which differ considerably from the corresponding DMC parameters) to derive
$E_\infty$ for a given $N$ from \Eref{eqfit2D} (DMC-SJ* and DMC-SJB*). 
}
\begin{indented}
\lineup
\item[]
\begin{tabular}{@{}l l l c  }
\br
\multicolumn{1}{c}{}&\multicolumn{1}{c}{}&\multicolumn{1}{c}{$E_\infty$ (Ha/e)}  & Reference\\
\mr
VMC-SJ    &  & 0.003886(5) & This work \\
VMC-SJB   &  & 0.002021(4) & This work\\
DMC-SJ    &  & 0.00301(1) & This work\\
DMC-SJB   &  & 0.001621(7) & This work\\
\mr
C-VMC &    &  0.002955&  \cite{Ceperley-1978}\\
CA-DMC-RN  &    &  0.002055& \cite{Ceperley-Alder-PRL-1980}\\
OB-VMC-SJ   &   &  0.0051(2)& \cite{Ortiz-Ballone-1994}  \\
OB-DMC-SJ  &    & 0.0033(2) & \cite{Ortiz-Ballone-1994}\\
KCM-VMC-SJ& &0.004096 &\cite{kwon98:_effec}$^{\rm a}$\\
KCM-DMC-SJ& & 0.002431& \cite{kwon98:_effec}$^{\rm a}$\\
KCM-DMC-SJB & & 0.001812& \cite{kwon98:_effec}$^{\rm a}$\\
\mr
DMC-SJ* & & 0.00266(1) & This work \\
DMC-SJB* & & 0.001380(8)& This work \\
\br
\end{tabular}
\item[]
$^{\rm a}$
These numbers for $r_s$=2 were obtained by using an updated form of Equation~(2)
 of \cite{Pitarke-2004}, which was originally derived by 
fitting three values of $r_s$ ($r_s$= 1, 5 and 10) of \cite{kwon98:_effec}
 and we have now fitted with the four available
 values of $r_s$ ($r_s$= 1, 5, 10 and 20) of \cite{kwon98:_effec} for each case (VMC-SJ, DMC-SJ and DMC-SJB).
 Differences between the energies derived from Equation~(2)
 of \cite{Pitarke-2004} and the numbers reported here for $r_s$=2
 are within $7\times10^{-5}$ Ha/e.
\end{indented}
\end{table}

\begin{figure}[htb]
\includegraphics*[width=0.90\linewidth]{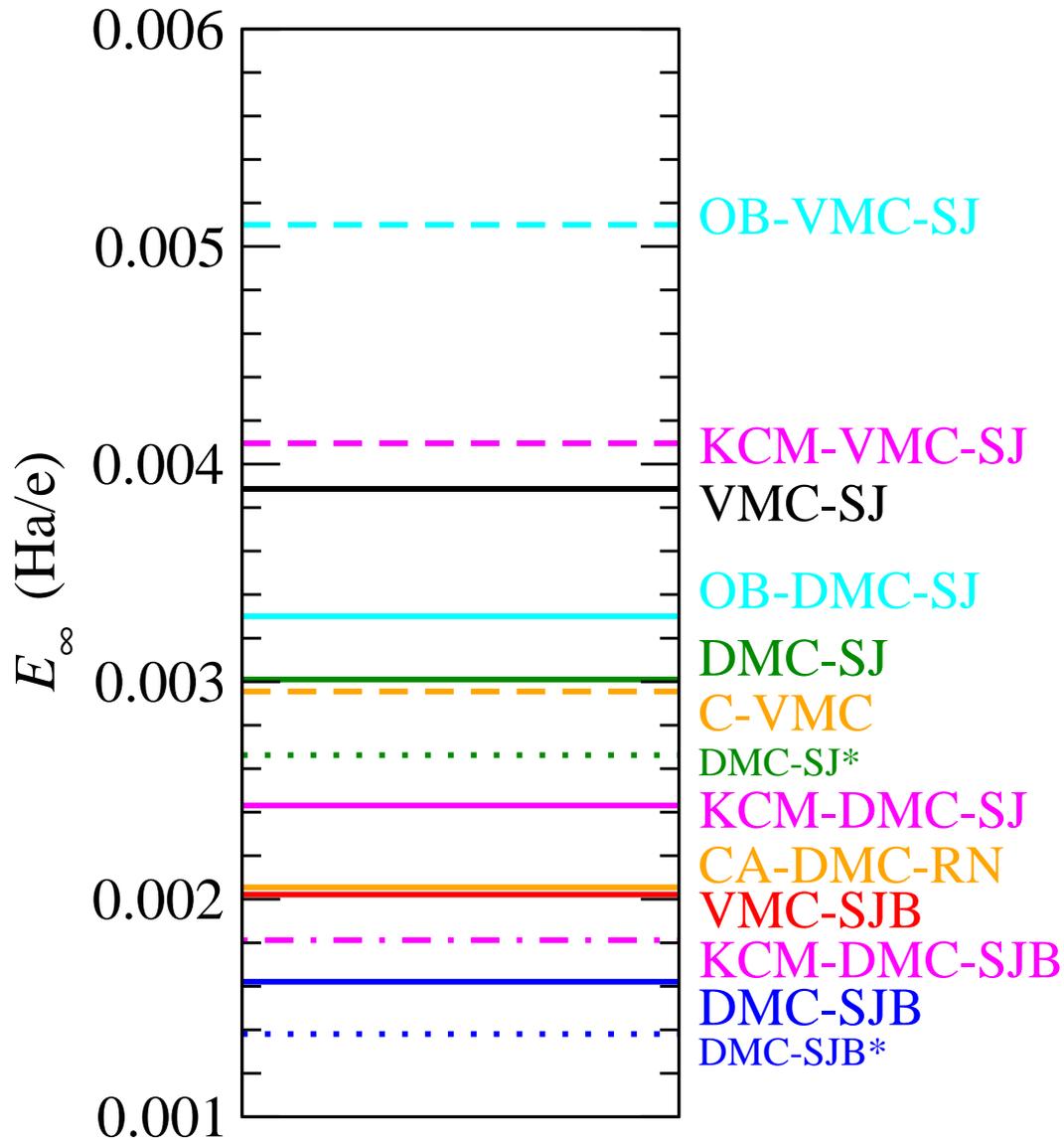}
\caption{\label{figEinf-comp}
(Color online) Schematic representation of the extrapolated energies quoted in table~\ref{tableEinf-comp}.
}
\end{figure}

Our extrapolated VMC and DMC ground-state total energies, $E_\infty$,
 are compared in table~\ref{tableEinf-comp} and \fref{figEinf-comp}
 to the corresponding energies reported in
\cite{Ceperley-1978, Ceperley-Alder-PRL-1980, Ortiz-Ballone-1994,kwon98:_effec}.
 These are: (i) the original VMC calculation of Ceperley \cite{Ceperley-1978},
 (ii) the released-node DMC calculation of Ceperley
 and Alder \cite{Ceperley-Alder-PRL-1980}, (iii) the fixed-node VMC and 
DMC calculations of Ortiz and Ballone \cite{Ortiz-Ballone-1994} and Kwon
 {\it et al.} \cite{kwon98:_effec}, 
and (iv) the backflow DMC calculation of Kwon
 {\it et al.} \cite{kwon98:_effec}. 
We note that the DMC calculations reported by Ortiz and Ballone
 (OB-DMC-SJ)\cite{Ortiz-Ballone-1994}
 and Kwon {\it et al.} (KCM-DMC-SJ and KCM-DMC-SJB) \cite{kwon98:_effec},
 were all carried out by first fitting within VMC the $E_\infty$, $b_0$, $b_1$,
 and $b_2$ parameters entering \Eref{eqfit2D} and \eref{eqfitOB} and then
 using the VMC parameters $b_0$, $b_1$, and $b_2$ to derive $E_\infty$ for
 a given $N$ from either \Eref{eqfit2D} or \eref{eqfitOB};
 hence, in order to compare to the results reported by these authors we
 have performed additional DMC fits (DMC-SJ* and DMC-SJB* represented
 by dotted lines in \fref{figEinf-comp}) by following this approximate
 extrapolation procedure. 
With this aim, we have calculated  $E_\infty$  from the DMC-SJ(B) data and the 
 $b_1$ and $b_2$ parameters of the VMC-SJ fit for each system size $N$, 
and then we have fitted a horizontal line to obtain the extrapolated DMC-SJ(B)* value 
reported in table \ref{figEinf-comp}.

 We find that (i) our DMC-SJ* calculation is very close to the corresponding
 calculation reported by Kwon {\it et al.} (KCM-DMC-SJ) \cite{kwon98:_effec},
 and (ii) our DMC-SJB* calculation is considerably lower than the 
corresponding KCM-DMC-SJB calculation, which is a signature of the
 better quality of our SJB trial wave functions.
We note that our calculations were performed using the MPC
interaction, while in the calculations reported in the above references
the Ewald interaction was used. Hence, 
we have repeated our DMC-SJ and DMC-SJB calculations using the usual $1/r$ 
Ewald interaction; these calculations
 (shown in table \ref{tableE-N} for DMC-SJ) indicate that in both cases the 
Ewald extrapolated energy is at most 
 0.3 mHa/e above the MPC extrapolated result,
which reflects the fact that after extrapolation to $N\to\infty$ the usual
 Ewald energy yields fairly good results.
We note, however, that the difference between the MPC and Ewald
DMC-SJ energy reported in  \tref{tableE-N} 
 for each N is significantly smaller than the finite-size correction proposed in
\cite{Chiesa-FSE} ($\Delta V = \omega_p/(4N)$, where $\omega_p$ is the plasmon energy).
This is because the MPC and Ewald energies in  \tref{tableE-N}  were obtained using 
different interaction schemes
in the branching part of the calculation, 
i. e., MPC and Ewald, respectively.
We have performed additional calculations of MPC energies using the Ewald interaction
 in the
branching factors, and we have found that only when the same interaction (Ewald) is used
in the branching part of the calculation, the difference between the MPC and Ewald energies agrees with the correction proposed by Chiesa {\it et. al} \cite{Chiesa-FSE}.

At this point, we focus our attention on a comparison between the DMC
 extrapolated values that we have obtained by
 (i) fitting within DMC the $E_\infty$,  $b_1$, and $b_2$
 parameters entering \Eref{eqfit2D} (DMC-SJ and DMC-SJB)
 and (ii) using the VMC parameters $b_1$ and $b_2$ to derive 
$E_\infty$ for a given $N$ from \Eref{eqfit2D} (DMC-SJ* and DMC-SJB*).
 Our calculations indicate that the size-dependences for VMC and DMC
 differ considerably;\footnote{Since the form and optimisation of the 
trial wave function entering VMC
calculations is size-dependent, in general, there is no reason to expect
 VMC and DMC finite-size extrapolations to be the same.} indeed, the DMC-SJ* and DMC-SJB* extrapolated values
 are too low, i.e., the use of VMC data to extrapolate the corresponding
 DMC calculations yields artificially lowered extrapolations. Hence,
 the KCM-DMC-SJB ground-state energy reported by 
Kwon {\it et al.} \cite{kwon98:_effec}
 nearly coincides with our more accurate DMC
 extrapolation (DMC-SJB) as a result of two 
competing effects: the KCM-DMC-SJB extrapolated energy is (i) higher than our better optimized DMC-SJB* calculation and (ii) too low due to the assumption (in the extrapolation procedure) that the VMC and DMC size dependences coincide.  

Finally, we note that although the fitting parameters $b_1$ and $b_2$ 
entering \Eref{eqfit2D} are not transferable from VMC to DMC calculations 
they are indeed transferable from DMC-SJ to the more expensive 
DMC-SJB calculations: The error introduced by using the DMC-SJ 
parameters $b_1$ and $b_2$ to derive the DMC-SJB $E_\infty$ 
for a given $N$ from \Eref{eqfit2D} is found to be of no 
more than 0.04(1) mHa/e.  
 
\subsection{Interaction energy} \label{Umpc}

\begin{figure}[htb]
\includegraphics*[width=0.90\linewidth]{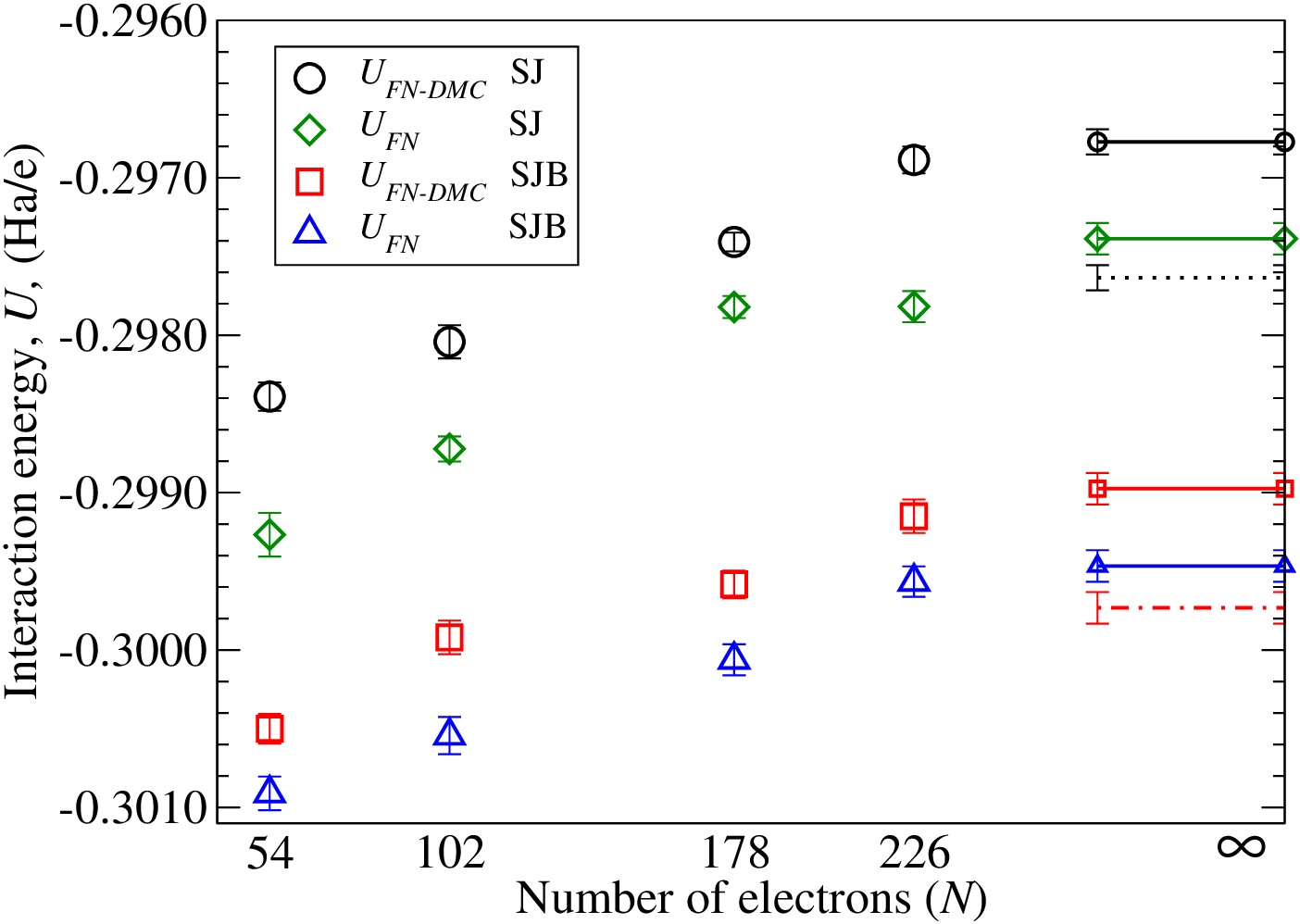}
\caption{\label{figU-N}
(Color online) Open symbols: The fixed-node interaction energies $U_{F\!N}$ and $U_{F\!N\!-\!D\!M\!C}$ quoted in table~\ref{table-U}. Open symbols joined by solid horizontal lines: The extrapolated interaction energies obtained from \Eref{eq-U-fit2D}.  
Black dotted and red dashed-dotted lines: The extrapolated estimator 2$U_{F\!N-D\!M\!C}-U_{V\!M\!C}$ for the infinite system, as obtained with the use of SJ (black dotted line) and SJB (red dashed-dotted line) trial wave functions.
}
\end{figure}

The interaction energy, $\hat U$, is a local operator (i.e., diagonal in real space) that
does not commute with the Hamiltonian. Hence, for an accurate calculation of the true expectation value of $\hat U$ we have applied the HF-based method of 
\cite{Gaudoin-Pitarke-HFS} described in \sref{HFSmethod}. 
\Fref{figU-N} and table~\ref{table-U} show the results that we have obtained for the true (HF) fixed-node interaction energies $U_{F\!N}=\langle\hat U\rangle_{H\!F}$ of a HEG of $r_s=2$ for $N$ = 54, 102, 178, and 226 electrons, 
as obtained by using either SJ or SJB trial wave functions. Also shown are the fixed-node interaction energies $U_{F\!N\!-\!D\!M\!C}=\langle\hat U\rangle_{F\!N\!-\!D\!M\!C}$ that we have obtained by using the standard DMC estimator, which are subject to an error that is linear in the difference between $\Psi_T$ and $\Psi_0^{F\!N}$.

\begin{table}[htb!]
\caption{\label{table-U}
Top: The fixed-node interaction energy of a HEG with $r_s$=2 for $N$= 54, 102, 178, and 226 electrons, as obtained by using the HF-based method of 
\cite{Gaudoin-Pitarke-HFS} ($U_{F\!N}$) and by using the standard DMC estimator ($U_{F\!N\!-\!D\!M\!C}$); both SJ and SJB trial wave functions have been used. The entry $N=\infty$ corresponds to the extrapolated values obtained 
from \Eref{eq-U-fit2D}. 
Bottom: The extrapolated estimator 2$U_{F\!N-D\!M\!C}-U_{V\!M\!C}$
for the infinite system. All energies are in Ha/e.
}
\begin{indented}
\item[]
\begin{tabular}{@{} l   l l l l l    }
\br 
 & \multicolumn{2}{c}{SJ wave function} & \multicolumn{2}{c}{SJB wave function} \\
$N$ & \multicolumn{1}{c}{$U_{F\!N}$ } &
\multicolumn{1}{c}{$U_{F\!N\!-\!D\!M\!C}$  } &
\multicolumn{1}{c}{$U_{F\!N}$ }  &
 \multicolumn{1}{c}{$U_{F\!N\!-\!D\!M\!C}$ }\\
\mr 
54 & -0.2993(1)& -0.29839(9) & -0.3009(1)& -0.30050(9) \\
102& -0.29872(8)& -0.2980(1) & -0.3005(1)& -0.2999(1)\\
178& -0.29782(7)& -0.29741(6) & -0.3001(1)& -0.29958(9)\\
226& -0.2978(1)& -0.29689(9) &  -0.2996(1)& -0.2991(1)\\
\mr
 $\infty$& -0.2973(1)& -0.29679(8) & -0.2994(1) & -0.2990(1) \\
\br
& & & & \\
\end{tabular}
\end{indented}

\begin{indented}
\item[]
\begin{tabular}{@{} c   l     }
\hline \hline
  &  \multicolumn{1}{c}{$2U_{D\!M\!C}-U_{V\!M\!C}$} (Ha/e)\\
\mr
SJ ($N=\infty$)&  -0.29764(8)\\
SJB($N=\infty$)&  -0.2997(1)\\
\br
\end{tabular}
\end{indented}
\end{table}

\begin{table}[htb!]
\caption{\label{table-UinfN}
The  fixed-node extrapolated interaction energy ($U_\infty$) for each system size
as obtained from the fits of \Eref{eq-U-fit2D}
of the  interaction energies of \tref{table-U}.
}
\begin{indented}
\item[]
\begin{tabular}{@{} l   l l l l l    }
\br
 & \multicolumn{2}{c}{SJ wave function} & \multicolumn{2}{c}{SJB wave function} \\
$N$ & \multicolumn{1}{c}{$U_{F\!N}$ } &
\multicolumn{1}{c}{$U_{F\!N\!-\!D\!M\!C}$  } &
\multicolumn{1}{c}{$U_{F\!N}$ }  &
 \multicolumn{1}{c}{$U_{F\!N\!-\!D\!M\!C}$ }\\
\mr
54 & -0.2970(2)& -0.2967(2) & -0.2993(2)& -0.2990(2) \\
102& -0.2975(1)& -0.2971(1) & -0.2997(1)& -0.2991(1)\\
178& -0.29714(9)& -0.29689(7) & -0.2996(1)& -0.2991(1)\\
226& -0.2973(1)& -0.29648(9) &  -0.2992(1)& -0.2988(1)\\
\br
\end{tabular}
\end{indented}
\end{table}

In order to obtain the extrapolated value $U_{\infty}$ from our finite-size calculations, we use the fitting equation
\begin{equation}\label{eq-U-fit2D}
U_N=U_\infty +\frac{b}{N} 
\end{equation}
for each set of data. 
The ${1}/{N}$ term does not aim at correcting the long-ranged errors,
  as we are using the MPC interaction.
It turns out, however, that the residual effects including shell-filling and the distortion of the xc hole due to the finite size geometry also seem to be well described by a 1/N fit.
The extrapolated values are displayed by 
the entry $N=\infty$ of table~\ref{table-U} and the 
solid lines of \fref{figU-N}, together with the result of
using the extrapolated estimator $2U_{F\!N\!-\!D\!M\!C}-U_{V\!M\!C}$ for the infinite system
(dotted and dashed-dotted lines).
 We see that while the true interaction energy $U_{F\!N}$ is overestimated by the standard DMC estimator
$U_{F\!N-D\!M\!C}$ (the error being linear in the difference between
$\Psi_T$ and $\Psi_0^{F\!N}$), it is underestimated by the extrapolated
estimator 2$U_{F\!N\!-\!D\!M\!C}-U_{V\!M\!C}$ (the error this time being quadratic in the difference between $\Psi_T$ and $\Psi_0^{F\!N}$). On the other hand, we see that as in the case of the ground-state total energy the effect of backflow (included in the calculations labeled SJB) is to lower the interaction energy by a rigid shift of about 2~mHa/$e$. 
Finally, we note that error bars in \fref{figU-N} 
are approximately the size of the symbols (about 0.1~mHa/e), even for the largest systems under consideration.
\Tref{table-UinfN} summarizes the extrapolated values
for each system size, $U_\infty =U_N-{b}/{N}$,
 for all the cases shown in \tref{table-U}.

In a recent paper \cite{Gaudoin-Pitarke-2007},
 it was demonstrated that accurate calculations of the interaction contribution to the ground-state energy of an arbitrary many-electron system can be obtained from the knowledge of the spherically averaged wavevector-dependent diagonal structure factor $S_k$ as follows
\begin{equation} \label{eq-SF-U}
U=\frac{1}{\pi}\int\left[S_k-1\right] \rmd k,
\end{equation}
where $S_k$ is the spherical average of the diagonal structure
factor in Fourier space:
\begin{equation}
S_k=1+\frac{4\pi}{N} \int \rmd{\mathbf r}n({\mathbf r})\int \rmd u \,u^2
\frac{\sin(ku)}{ku} n_{xc}({\mathbf r},u).
\end{equation}
Here, $N$ is the particle number, 
$n({\mathbf r})$ is the electron density at {\bf r}, and
$n_{xc}({\mathbf r},u)$ is the spherically averaged exchange-correlation hole
density $n_{xc}({\mathbf r},{\mathbf r'})$ at ${\mathbf r'}$ around an electron 
at {\bf r}.
If one samples the structure factor using only correlations within the 
simulation cell, then \Eref{eq-SF-U} represents the $k$-resolved MPC 
interaction \cite{Gaudoin_etal-SFU-dmc-2009}. 

\begin{figure}[htb]
\includegraphics*[width=0.90\linewidth]{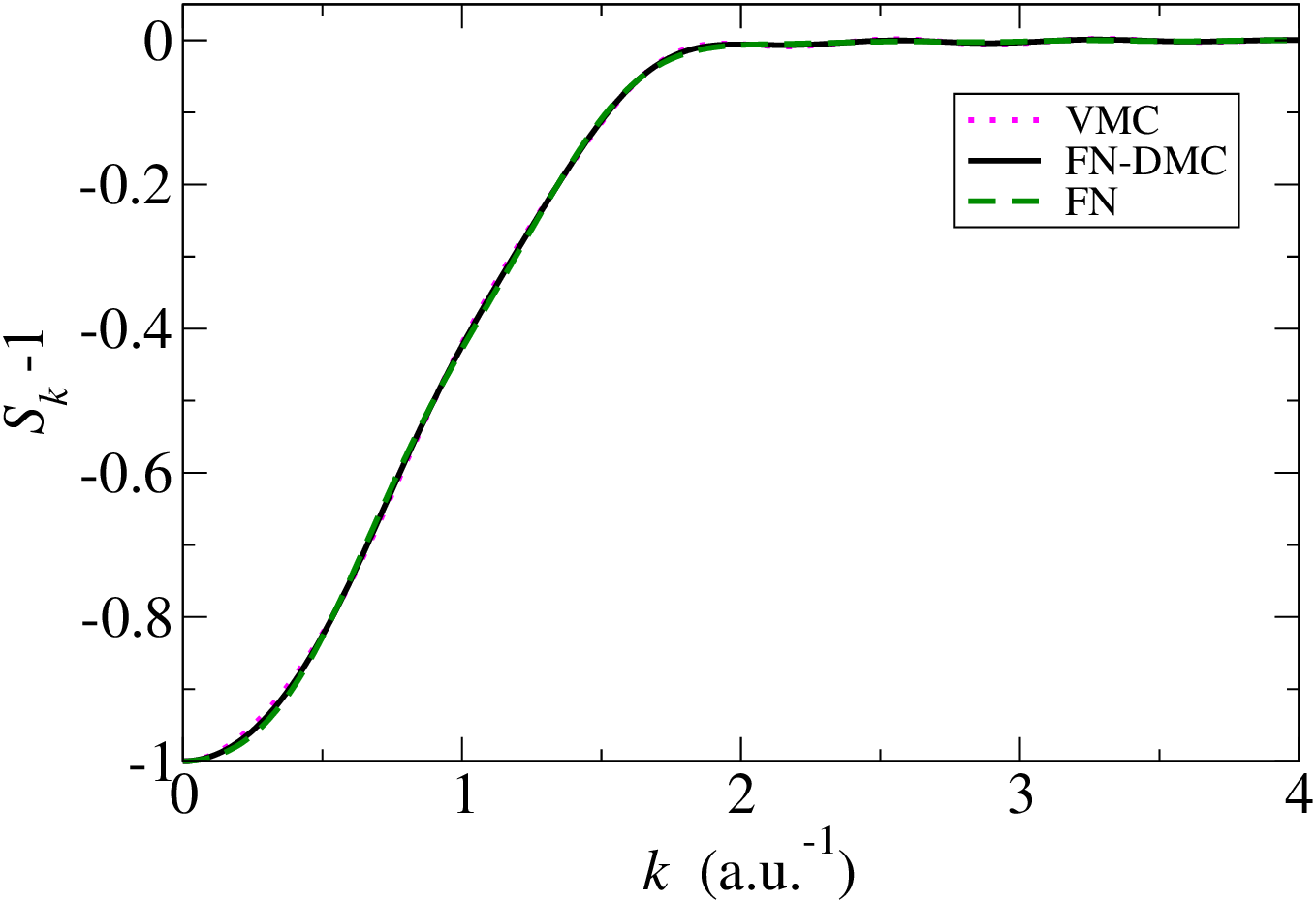}
\caption{\label{fig-SFk}
(Color online) VMC, standard-DMC, and HF-based DMC calculations of the spherically averaged structure factor $S_k$ of a HEG with $r_s$=2 and $N$=102 electrons, as obtained with
the use of SJ trial wave functions. 
}
\end{figure}

The structure factor can be computed using VMC \cite{Gaudoin-Pitarke-2007},
 standard DMC,
or the HF-based DMC \cite{Gaudoin_etal-SFU-dmc-2009}.
Since the spherically averaged structure factor is a diagonal function in real space, HF-based DMC calculations should yield the exact fixed-node $S_k$.
\Fref{fig-SFk} exhibits VMC, standard-DMC  (FN-DMC) and HF-based DMC (FN)
 calculations of $S_k$ for a HEG with $r_s=2$ and $N$=102 electrons,
 as obtained with the use of SJ wave functions.
\Eref{eq-SF-U} then yields the following VMC, standard-DMC, and HF-based
DMC interaction energies:
 -0.29717(3)~Ha/e, -0.2979(1)~Ha/e, and -0.29871(8)~Ha/e,
 respectively, which agree with our 
calculated expectation values $\langle\hat U\rangle_{V\!M\!C}=-0.29719(3)$~Ha/e,
$\langle\hat U\rangle_{F\!N\!-\!D\!M\!C}=-0.2980(1)$~Ha/e, and
$\langle\hat U\rangle_{H\!F}=-0.29872(8)$~Ha/e (see also table~\ref{table-U}), as expected.

\subsection{Kinetic energy} \label{KE}

\begin{table}[ht!]
\caption{\label{table-delta}
Standard-DMC (top) and HF-based DMC (bottom) kinetic and interaction energies, as obtained in the thermodinamic limit ($N\to\infty$) with the use of SJ and SJB wave functions.
$\Delta=|SJ-SJB|$ denotes the absolute value of the difference between the SJ and SJB
 calculations.}
\begin{indented}
\item[]
\begin{tabular}{@{}l l l }
\br
\multicolumn{1}{c}{}&\multicolumn{1}{c}{$U_\infty$ (Ha/e)}&\multicolumn{1}{c}{T$_\infty $(Ha/e)} \\
\mr
FN-DMC SJ    & -0.29679(8) & 0.29980(8) \\
FN-DMC SJB   & -0.2990(1) & 0.3006(1) \\
$\Delta$=$|$SJ-SJB$|$   & 22(1)10$^{-4}$  & 8(1)10$^{-4}$   \\
\br
FN SJ    & -0.2973(1) & 0.3003(1) \\
FN SJB   & -0.2994(1) & 0.3011(1) \\
$\Delta$=$|$SJ-SJB$|$   &  22(1)10$^{-4}$ & 8(1)10$^{-4}$  \\
\br 
\end{tabular}
\end{indented}
\end{table}

Assuming that the ground-state interaction and total energies have been correctly extrapolated, 
the kinetic energy can be obtained in the thermodynamic limit as
$T_{\infty}= E_{\infty}-U_{\infty}$. Table~\ref{table-delta} shows
standard-DMC and the more accurate HF-based DMC calculations of $T_{\infty}$,
 as obtained from the corresponding standard-DMC and HF-based DMC
 calculations of $U_{\infty}$ (also quoted) with the use of either
 SJ or SJB trial wave functions. The absolute value of the difference between
 the SJ and SJB calculations ($\Delta=|{\rm SJ}-{\rm SJB}|$) is also shown
 in this table. We note that this difference is considerably larger 
for the interaction energy, both in the case of the standard DMC
 approach and in the case of the more accurate HF-based DMC approach.
This is an indication of the fixed-node error being smaller in the kinetic energy than in the interaction energy.
Indeed, the correlation contribution to the kinetic energy is always smaller than the
corresponding  contribution to the interaction energy.

\section{Conclusions}

We have presented benchmark VMC and DMC ground-state energies of a 3D HEG with $r_s=2$ and
$N$=54, 102, 178, and 226 electrons, using an MPC interaction and backflow corrections. We have extrapolated our finite-size calculations to the thermodynamic limit, and we have found lower energies than previously reported, thus showing the good quality of our fixed-node trial wave functions. We have shown that previously extrapolated results are artificially lowered by assuming that the VMC and DMC size dependences (which we analyze independently) coincide.
We have used the HF operator sampling method introduced in 
\cite{Gaudoin-Pitarke-HFS} to compute accurate values of the kinetic and interaction contributions to the ground-state energy. We also show that these values, as obtained with the use of the MPC interaction, coincide with the result one obtains from the spherically averaged structure factor. Our calculations indicate that our HF-based DMC approach yields very accurate results even for very large systems.
Finally, we have found that the difference between the interaction
 energies that we obtain using either the original Slater-determinant
 nodes or the backflow-displaced nodes is considerably larger
 than the difference between the corresponding kinetic energies.
A combination of (i) the fact that our Hellman-Feynman operator sampling 
method allows, within the fixed-node approximation, to calculate accurately 
the kinetic-energy contribution to the ground-state energy and 
(ii) the fact that the fixed-node error is smaller in the kinetic 
energy than in the interaction and total ground-state energy leads
 us to the conclusion that our kinetic energies should be of great
 use in the construction of accurate kinetic-energy functionals.

\ack{
This work was partially supported by the Basque
Government (Grant No. GIC07IT36607) and the Spanish Ministerio de Educaci\'{o}n y Ciencia (Grant. No. FIS2006-01343 and CSD2006-53).
Computing facilities were provided by the Donostia International Physics Center (DIPC) and
the SGI/IZO-SGIker UPV/EHU (supported by the National Program for the Promotion of Human
Resources within the National Plan of Scientific Research, Development
and Innovation-Fondo Social Europeo, MCyT and the Basque Government).
}

\section*{References}
\bibliographystyle{unsrt}
\bibliography{qmc}

\end{document}